%% file: main.tex
\documentclass[a4paper,11pt]{article}
\usepackage{aaskaiid,orcidlink}
\usepackage{xcolor,graphicx}
\newcommand{\Rp}{\ensuremath{R_{\text{pl}}}}

\newcommand{\Reff}{\ensuremath{R_{\text{eff}}}}
\newcommand{\Rmp}{\ensuremath{R_{\text{mp}}}}

\newcommand{\vw}{\ensuremath{v_{\text{w}}}}

\title{Radio emission from star-planet interactions}
\ShortTitle{Radio SPI}

\author[1,2]{H.K. Vedantham\orcidlink{0000-0002-0872-181X}}
\ShortName{Vedantham et al.} 
\author[3]{A. Strugarek\orcidlink{0000-0002-9630-6463}}
\author[4]{C.K. Louis\orcidlink{0000-0002-9552-8822}}
\author[1,5]{J.R. Callingham\orcidlink{0000-0002-7167-1819}}
\author[6]{L. Peña-Moñino\orcidlink{0000-0001-6735-1655}}
\author[6]{M. P\'erez-Torres\orcidlink{0000-0001-5654-0266}}
\author[4,7]{P. Zarka\orcidlink{0000-0003-1672-9878}}
\author[4]{E. Mauduit\orcidlink{0000-0001-5294-2890}}
\author[4]{N. Duchêne\orcidlink{0000-0003-3224-5603}}
\author[8]{P. Amado\orcidlink{0000-0002-8388-6040}}
\author[1,2]{S. Bloot\orcidlink{0000-0002-3601-6165}}
\author[5,1]{R.~D.~Kavanagh\orcidlink{0000-0002-1486-7188}}
\author[9]{A. A. Vidotto\orcidlink{0000-0001-5371-2675}}
\author[4,10]{L. Lamy\orcidlink{0000-0002-8428-1369}}
\author[4,7,11]{C. Tasse\orcidlink{0009-0009-9030-7885}}

\affiliation[1]{ASTRON, Netherlands Institute for Radio Astronomy, Oude Hoogeveensedijk 4, 7991\,PD, Dwingeloo, The Netherlands}
\emailAdd{vedantham@astron.nl}

\affiliation[2]{Kapteyn Astronomical Institute, University of Groningen, Landleven 12, 9747\,AD Groningen, Netherlands}

\affiliation[3]{Département d’Astrophysique/AIM, CEA/IRFU, CNRS/INSU, Université Paris-Saclay, Université de Paris}

\affiliation[4]{LIRA, Observatoire de Paris-PSL, CNRS, Sorbonne Université, Université Paris Cité, 92190 Meudon, France}

\affiliation[5]{Anton Pannekoek Institute for Astronomy, University of Amsterdam, Science Park 904, 1098\,XH Amsterdam Nederland}

\affiliation[6]{Institut de Ci\'encies de l’Espai (ICE, CSIC), Campus UAB, Carrer de Can Magrans s/n, 08193 Bellaterra, Spain}

\affiliation[7]{ORN, Observatoire de Paris-PSL, CNRS, Université d'Orléans, 18330 Nançay, France}

\affiliation[8]{Instituto de Astrof\'isica de Andaluc\'ia, Glorieta de la Astronom\'ia, Spain}

\affiliation[9]{Leiden Observatory, Leiden University, P.O. Box 9513, 2300 RA Leiden, The Netherlands}

\affiliation[10]{LAM, Aix Marseille Universit\'e, CNRS, CNES, Marseille, France}

\affiliation[11]{Centre for Radio Astronomy Techniques and Technologies (RATT), Department of Physics and Electronics, Rhodes University, Makhanda, 6140, South Africa.}

\abstract{Stars interact with their exoplanets though gravity, radiation, plasma and magnetic fields. Stellar plasma and magnetic fields impose electrodynamic effects on exoplanet atmospheres and interiors that include heating, aurorae and atmospheric mass loss. The planets in turn can excite Alfv\'enic disturbances that are dissipated on the star leading to chromospheric heating and flares. This interaction, broadly called magnetic star--planet interaction (M-SPI), can also generate radio signatures both from the star and the exoplanet. The radio emission encodes information on the dynamics/energetics of the interaction, the magnetic field strength and topology of the emitter and the orbital/rotational geometry of the system--- information that is difficult or in some cases implausible to obtain by other means. Yet we do not have a conclusive detection of M-SPI in the radio band primarily due to sensitivity limitations and scarce observing time spent monitoring promising targets. Here we describe the scientific motivation to study M-SPI in exoplanetary systems, to get progress in understanding its predicted signal strength and phenomenology. We argue that the SKA telescopes can make a transformative contribution to exoplanet science by detecting M-SPI in the radio band but this will require substantial observing time--- similar to that afforded to successful optical-band searches for M-SPI signatures.}

\begin{document}
\maketitle

\section{Scientific Motivation}
\label{sec:intro}
The space weather environment of a star plays a pivotal role in the atmospheric evolution of exoplanets. 
The stellar wind and coronal mass ejections can strip planetary atmospheres over gigayear timescales \citep{2007AsBio...7..185L,2020JGRA..12527639G}. 
In addition, fluctuations in the interplanetary magnetic field of the stellar wind can induce eddy currents that dissipate heat in planetary ionospheres, further exacerbating atmospheric loss \citep{2025A&A...693A.220S}.
Conversely, close-in exoplanets can tug on stellar magnetic field lines, launching Alfv\'en waves that transport energy back to the stellar coronae and chromospheres \citep{2004ApJ...602L..53I,2007P&SS...55..598Z,2013A&A...552A.119S}. 
Such star--planet interactions, mediated by magnetised plasma, are generally referred to as \textit{magnetic star--planet interactions} (hereafter M-SPI), or \textit{wind--magnetosphere interactions} to distinguish them from interactions mediated by gravity or radiation.

The magnitude of M-SPI's potentially detrimental effects on exoplanetary atmospheres depends on stellar wind conditions, the detailed physics of the interaction, and the presence and strength of a planetary magnetic field. 
Magnetohydrodynamic (MHD) simulations can be used to study these effects in various settings \citep[e.g. ][]{2021MNRAS.504.1511K,2022MNRAS.512.4556S}, but their application to specific systems is limited by the lack of empirical data on stellar wind ram pressure, magnetic pressure at the planet’s location, and the magnetic field strength of the exoplanet itself. 
Furthermore, non-ideal MHD effects such as particle acceleration, charge exchange, and wave--particle interactions are difficult to model from first principles. 
Empirical data on the occurrence and energetics of M-SPI are therefore essential for advancing the field.
Observational efforts to detect M-SPI signatures have intensified in recent years and recent results in the optical wave-band demonstrate that such interactions exist and can provide empirical estimates of the interaction energetics \citep[][;Revilla et al., Science, accepted]{2018AJ....156..262C,2019NatAs...3.1128C,2022MNRAS.512.5067K,2025Natur.643..645I}. 
These efforts underscore the promise of detecting and studying radio signatures that can provide complementary and unique information \citep{2024NatAs...8.1359C}. 
In this chapter, we describe the transformative potential of Square Kilometre Array (SKA) telescopes to study M-SPI from nearby ($d\lesssim 10^2\,{\rm pc}$) stellar/planetary systems. We also refer the reader to the accompanying chapter by \citet{Kavanagh01.2026.SKA} that focusses on direct magnetospheric emission from exoplanets and the chapter on such emission by \citet{2015aska.confE.120Z} in the preceding SKA science book \citep{2015aska.confE.174B}.

\subsection{Sub- and Super-Alfv\'enic regimes}
\label{subsec:sub_super_spi}
\begin{figure}
    \centering
    \includegraphics[width=\linewidth]{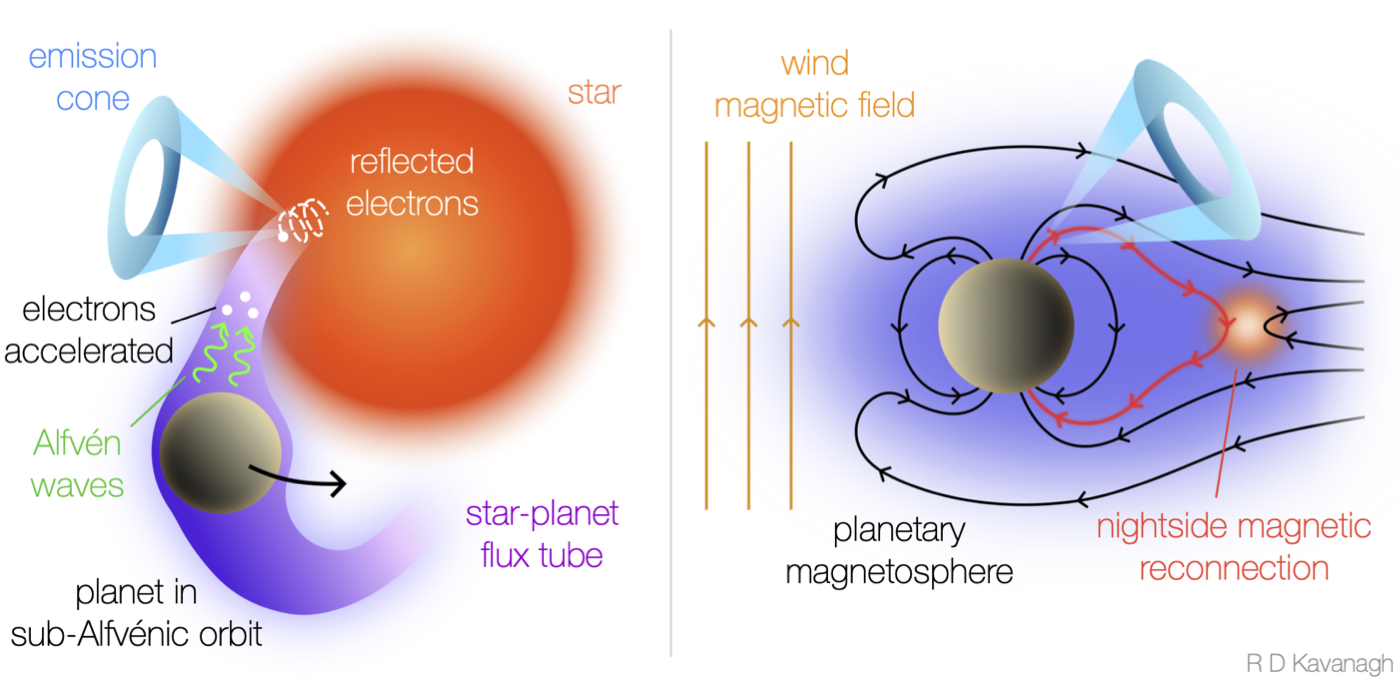}
    \caption{Sketch showing M-SPI regimes. Left panel: Sub-Alfv\'enic case where Alfv\'en wing/s (purple tube) created by the planet's flow connect back to the star and produce radio emission (cyan cone) in the stellar magnetosphere. Part of the energy can also produce radio emission in the magnetosphere of the planet (not shown in figure).  Right panel: Super Alfv\'enic case where energy flow back to the star is not possible but the intercepted energy can power radio emission in the planet's magnetosphere (cyan cone). Figure reproduced from \citet{2024NatAs...8.1359C}.}
    \label{fig:spi_sketch}
\end{figure}

M-SPI can be conceptualised as a planet acting as an obstacle to the magnetised plasma flow from the host star. 
The consequences of this interaction differ significantly depending on whether the plasma flow speed, in the planet’s frame of reference, is above or below the speed at which magnetic disturbances propagate—the so-called Alfv\'en speed (see Fig. \ref{fig:spi_sketch}).
In the \textit{sub-Alfv\'enic} regime (flow speed $<$ Alfv\'en speed), the plasma has sufficient time to respond to the interruption caused by the planetary obstacle. 
In this case, a portion of the intercepted energy is transported back to the star via Alfv\'en waves, while the remaining energy is dissipated locally in the planet’s atmosphere or magnetosphere. 
A fraction of the energy carried back to the star is ultimately dissipated in the stellar corona and/or chromosphere. The benchmark system for sub-Alfv\'enic interaction is the Jovian interaction with the Galilean moons \citep{1969ApJ...156...59G,1998JGR...10319843N,2018A&A...618A..84Z}.

In contrast, in the \textit{super-Alfv\'enic} regime, the plasma cannot respond rapidly enough to the obstacle—analogous to the formation of a bow shock in front of a supersonic aircraft. 
The result is the formation of a bow shock upstream of the planetary obstacle and a contact discontinuity that separates the stellar wind from the planetary atmosphere or magnetosphere. 
In this regime, upstream energy transport is suppressed, and the interaction energy is dissipated in the vicinity of the planet and carried away downstream.

In the literature, the sub-Alfv\'enic regime is sometimes implied by M-SPI and the super-Alfv\'enic regime is referred to as `wind--planet' or `wind--magnetosphere' interaction. Regardless, in both regimes, a substantial fraction of the wind energy intercepted by the obstacle can be converted into electromagnetic radiation \citep{2007P&SS...55..598Z,2013A&A...552A.119S}. 
This emission can be detected from Earth and used to probe the physical properties of the interaction. 
We now describe these electromagnetic signatures, with a particular focus on radio emission detectable by the SKA telescopes.

\subsection{Signatures of magnetic SPI}
\label{subsec:spi_signatures}
M-SPIs have been searched for at different wavelength regimes. 
In general, the hunt consists in finding a signal modulated at the known exoplanet orbital period. 
Evidence for M-SPI has been found in some stellar activity tracers of the host star, especially in the H and K lines of Ca II \citep{2003ApJ...597.1092S,2019NatAs...3.1128C}. 
For these systems, once the stellar rotational modulation is removed, a signal modulated at the hot-Jupiter orbital period is unveiled. 
Nevertheless, some of these detections were successful only at specific observational periods (e.g. one out of 7 epochs for HD 189733, \citep{2018AJ....156..262C}, which shows an on/off mechanism for the existence of this particular periodic signal in the activity tracer \citep{2008ApJ...676..628S}. 
Similar exoplanet-phase modulated signals were also detected in X-ray \citep[e.g.,][]{2015ApJ...811L...2M, 2023ApJ...951..152A} and far UV spectroscopy \citep{2015ApJ...805...52P}. 
Broad-band photometry has also been used to detect hints of M-SPI. 
Early studies with MOST \citep{2008A&A...482..691W} and CoroT \citep{2009EM&P..105..373P} found tentative modulations of the photometric signals associated with a known close-in exoplanet. 
\citet{2024A&A...684A.160C} detected a temporally-variable signal in TESS for HD 118203 which can be explained convincingly so far only with M-SPI. 
Two results from long-term optical monitoring have provided compelling statistical evidence for the existence of M-SPI. 
The first employed TESS and CHEOPS photometry to detect a statistical excess of flares in HIP 67522 near the transit phase of its close-in exoplanet, pointing towards the inducement of stellar flares by the close-in exoplanet in the system \citep{2025Natur.643..645I}.  
The second employed nearly two decades of optical spectroscopy of GJ\,436 with HARPS and CARMENES to reveal variability in the chromospheric Ca II H\&K and Ca II infrared triplet lines at its planet’s synodic (2.81 d) and sum (2.46 d) periods \citep{Revilla2026}. 
The measurement of the energetics of the M-SPI signatures in these two cases have been used to estimate the atmospheric erosion rate (in case of HIP 67522) and the plausible range of magnetic field strength of the interacting exoplanet (in case of GJ\,436), demonstrating the unique diagnostic power of M-SPI observations.

All the aforementioned signatures are \textit{indirect} tracers of the existence of M-SPI as similar signatures could in principle also be triggered from tidal effects. 
As a result, their interpretation as M-SPI rely on the use of models to deduce physical quantities from their detection \citep[see e.g.,][]{2019NatAs...3.1128C,2022MNRAS.512.4556S}. 
The signature of M-SPI in the radio band is expected to be the so-called electron cyclotron maser instability (ECMI hereafter) emission. This is a coherent emission mechanism that occurs close to the local cyclotron frequency, given by $\nu_{\rm c} = 2.8\times(B/{\rm Gauss})\,{\rm MHz}$ where $B$ is the magnetic field strength at the emitter \citep{2006A&ARv..13..229T}. 
The emission is highly beamed along the surface of a cone whose axis is parallel to the magnetic field line \citep{2006A&ARv..13..229T}. 
In case of sub-Alfv\'enic interaction, the ECMI emission could originate in the magnetosphere of the planet (if it is magnetized) and/or in the magnetosphere of the star. 
In the super-Alfv\'enic case, since energy transport back to the star is suppressed, ECMI only appears from the magnetosphere of the planet (again, provided it is magnetised).

The maximum frequency of ECMI emission from a magnetosphere is given by the cyclotron frequency at the surface of the emitting body. 
The resulting cut-off of the ECMI radio spectrum provides a direct, model-independent measurement of the magnetic field of the emitter. 
For example, Jupiter's ECMI emission is restricted to frequencies below 40\,MHz. 
Canonical planetary dynamo scaling laws predict stronger field strengths for gas-giants with faster rotation rates, larger masses and heat flux \citep{2010SSRv..152..565C} which will push the emission of several gas-giant exoplanets into the SKA-Low band; however, see also \citet{2018ApJ...862...19Z}. 
Indeed, a measurement of exoplanets' magnetic field via detection of M-SPI emission from the planet is a high-impact and long standing goal of radio astronomy. 
Exoplanet-side emission and its detection prospects with the SKA telescopes are described in the companion chapter by \citet{Kavanagh01.2026.SKA}. Here we primarily focus on star-side ECMI emission but make cursory remarks on the exoplanet-side emission given the common causes and energetics of the two emission components.

For the star-side emission, suppression of the ECMI mechanism at frequencies below the ambient plasma frequency \citep{2006A&ARv..13..229T} will likely preclude detection of M-SPI from Sun-like stars that have weak large scale fields. 
However, the most common stars, M-dwarfs are expected to have fields strengths well in excess of tens of Gauss \citep{morin10, lehmann24}, which will push their surface cyclotron frequencies above typical coronal plasma frequencies. 
The cyclotron frequencies are also well into the SKA-Low band and in many cases, even the SKA-mid bands, which makes the SKA telescopes suitable for their detection. 

So far, only tentative, and often disputed, radio detections of M-SPI signatures have been published (e.g. \citealt{2021A&A...645A..59T,2023NatAs...7..569P,2025A&A...700A.140Z,2020NatAs...4..577V,2021A&A...645A..77P,2013A&A...552A..65L,2014A&A...562A.108S} with numerous non-detection \citep{2018MNRAS.478.2835L,2025A&A...693A.162C,2015MNRAS.446.2560M,2017MNRAS.467.3447L,2010AJ....139...96L,2013ApJ...762...34H,2000ApJ...545.1058B,2023ApJ...952..118R,2024A&A...682A.170B}.
The contested detections usually argue in favour of M-SPI based on the implausibility of the emission originating by other means (inactive star; e.g. \citet{2020NatAs...4..577V}) or an association of radio bursts with particular orbital phases of a known exoplanet (e.g. \citet{2023NatAs...7..569P}). More recently, \citet{Tasse2026} have argued for M-SPI radio emission in GJ\,687 based on `arch-like' morphology that is expected in case of M-SPI due to beaming-related visibility at different frequencies.

The likely reasons for the lack of an uncontested/secure radio M-SPI detection are (a) confusion from unrelated stellar ECMI emission, (b) the relatively low amount of radio telescope time spent precluding a robust statistical detection of the tell-tale periodicity of M-SPI, (c) sensitivity limitations of current telescopes and (d) the highly beamed nature of ECMI \citep{2006A&ARv..13..229T} which means that a favourable viewing geometry is necessary for detection \citep{2011A&A...531A..29H,2023MNRAS.524.6267K} regardless of telescope sensitivity.
 
For most star--planet systems, we do not know a priori if the radio beam crosses the Earth, making target selection difficult. 
Moreover, without independent knowledge of the plasma densities in the ECMI emission region, it is not possible to identify targets where the emission is not catastrophically suppressed and absorbed in the vicinity of the source itself \citep{2018MNRAS.479.1194D,2017A&A...602A..39V}. 
Therefore, a pragmatic strategy for M-SPI search in the radio would be to combine blind surveys to detect the tell-tale signs of ECMI with targeted surveys of a handful of systems where substantial auxiliary data on the stellar magnetic field and plasma densities is available that point to (a) non-suppression of ECMI emission and/or (b) a favourable viewing geometry. 
We discuss observational strategy to maximum chances of detection in \S\ref{sec:obs} and first consider the question of anticipated radio flux densities for comparison with the SKA telescopes' target sensitivity values. 

\section{Radio flux density predictions}
\label{sec:scaling-laws}

\subsection{Scaling law flavours}
\label{subsec:scaling_laws}
The energy that powers the observed radio emission at both the star- and exoplanet-ends scales with the amount of Poynting flux in the stellar wind intercepted by the planetary obstacle. 
There exist several scaling laws that postulate different conversion efficiencies between the intercepted Poynting flux and the emitted radio power, such as the Alfv\'en wing model (e.g. \citealt{2007P&SS...55..598Z,2013A&A...552A.119S,2016ApJ...833..140S}), the reconnection model \citep{2009A&A...505..339L} and the stretch-and-break model \citep{2013A&A...557A..31L,2022MNRAS.512.4556S}.
In the Alfvén wing model, a fraction of the intercepted Poynting flux propagates back to the host star via one of the Alfv\'en wings. 
The acceleration of charges that emit the radio-waves is expected to occur on kinetic scales in the Alfv\'en wings \citep{2007JGRA..11211212H}. 
In the reconnection model, particle acceleration happens at the planetary magnetospheric boundary due to continuous reconnecting between the planetary magnetic field and the  stellar-wind magnetic field. 
In the stretch-and-break model, reconnection is induced by the orbital motion of the planet, which twists magnetic field lines and triggers stronger reconnection events.
Finally, an empirical scaling law has also been postulated based on the observed radio power of the magnetised planets in the solar system as well as the radio power of the emission induced on Jupiter due to sub-Alfv\'enic interaction with its larger moons, notably Io \citep{2007P&SS...55..598Z, Zarka2025}.  

In all of the above models, the final result is a release of magnetic energy but the models posit different efficiencies. 
The key quantity that sets the energetics in all models is the intercepted Poynting flux given by $S_{\rm Poynt}
\propto \, (B_{\rm sw}\, R_{\rm eff})^2\,v_{\rm rel}$, where 
 $v_{\rm rel}$ is the
relative velocity between that of the stellar wind flow, $\vw$, and that of the planet; $B_{\rm sw}$ is the
stellar wind magnetic field perpendicular at
the location of the planet; and  
$R_{\rm eff}$ is the effective radius of the exoplanet.  
In the Alfvén wing model, the planet may be magnetized, or unmagnetized, as long as the (unmagnetized) planet is an electric conductor. Therefore, $\Reff \geq \Rp$. 
In the reconnection and stretch-and-break scenarios, not only must the planet be magnetized for the model to work, but the intensity of the magnetic field needs to be large enough, so that a planetary magnetosphere is formed. 

The fractional Poynting flux that can potentially be converted to radiation in the Alfvén wing model is given by \citep{2007P&SS...55..598Z,2013A&A...552A.119S}:
\begin{equation}
    S_{\rm Alf} = \frac{\overline{\alpha}\,M_{\rm A}}{2}(B_\perp\,\Reff)^2\,v_{\rm rel}, 
\end{equation}
where $B_\perp$ is the component of $B_{\rm sw}$ that is perpendicular to the wind velocity in the frame of the planet, $M_{\rm A}$ is the Alfvén Mach number, $v_{\rm A}$ is the Alfvén speed, and $\overline{\alpha}$ 
is a measure of how effective the interaction is. Here and throughout, we use Gaussian units unless specified otherwise.

For the reconnection model, the expression is given by \citep{2009A&A...505..339L}:
\begin{equation}
\label{eq:Lanza_eq8}
S_{\rm rec} = \frac{\gamma}{4} (B_{\rm sw} \Rmp)^2\,v_{\rm rel}, 
\end{equation}
where   $0 < \gamma  <1$, and its exact value depends on the angle between the interacting magnetic field lines, reaching its maximum when the magnetic fields of the planet and the star are anti-aligned.

Finally, for the stretch-and-break model we have (e.g. \citealt{2013A&A...557A..31L,2022MNRAS.512.4556S}):
\begin{equation}
\label{eq:stretch_and_break}
    S_{\text{sb}} = \frac{f_{\rm ap}}{2} (B_{\rm p} \Rmp)^2 v_{\text{rel}}
\end{equation}
where $B_{\rm p}$ is the planetary magnetic field strength,  and $f_{\text{ap}}$ is the fractional area of the planetary disc where magnetic field lines are connected to the stellar wind. It is given by 
\begin{equation}
    f_{\rm ap} = 1-\left(1- \frac{3\xi^{1/3}}{2+\xi}\right)^{1/2};\,\,\xi = \frac{B_{\rm sw}}{B_{\rm p}}
\end{equation}
These different scaling laws can lead to estimates differing by one to three orders of magnitudes depending on the parameters \citep{2022MNRAS.512.4556S}, requiring a more robust and thorough empirical law. Analysing 3D global numerical simulations of the 3D M-SPI, \citet{2025arXiv251023277P} assessed the maximal Poynting flux converted by the planetary obstacle and obtained a coherent upper limit such that
\begin{equation}
    \label{eq:max}
    S_{\rm max} = 5.382 \left(R_{\rm p}^2 B_{\rm sw}^2 v_{\rm rel}\right) \left(\frac{B_{\rm p}}{1{\rm G}}\right)^{0.5}\left(\frac{1{\rm G}}{B_{\rm sw}}\right)^{0.85} \, [{\rm erg/s}] \, .
\end{equation}
This upper limit was obtained considering the Poynting flux going towards the star only. It gives values on par, for reasonable parameters, with the high estimates of the stretch-and-break model ($S_{\rm sb}$). 

The total power liberated in the radio-band may be written as $P_R = \beta\,S_{\rm Poynt}$, where $\beta$ is the efficiency factor in converting Poynting
flux to ECM radio emission. 
Although the value of $\beta$ cannot be accurately determined from first-principles in the above models, based on empirical scaling seen in the solar system, it is expected to be 
in the range from $10^{-4}$ to $10^{-2}$, with a widely-adopted nominal value of 
$\beta = 2\times 10^{-3}$ \citep{2007P&SS...55..598Z,Zarka2025}.
Finally, the radio flux density observed on Earth is

 \begin{equation}
   \label{eq:flux_density}
   F_R = \frac{P_R}{\Omega\,D^{2} \,\Delta\nu}, 
 \end{equation}

where $\Omega$ is the beam-solid angle of the emission, $D$ is the distance to the emitter and $\Delta \nu$ is the emission bandwidth. 
\begin{figure}
\begin{minipage}{0.65\textwidth}
\includegraphics[width=\linewidth]{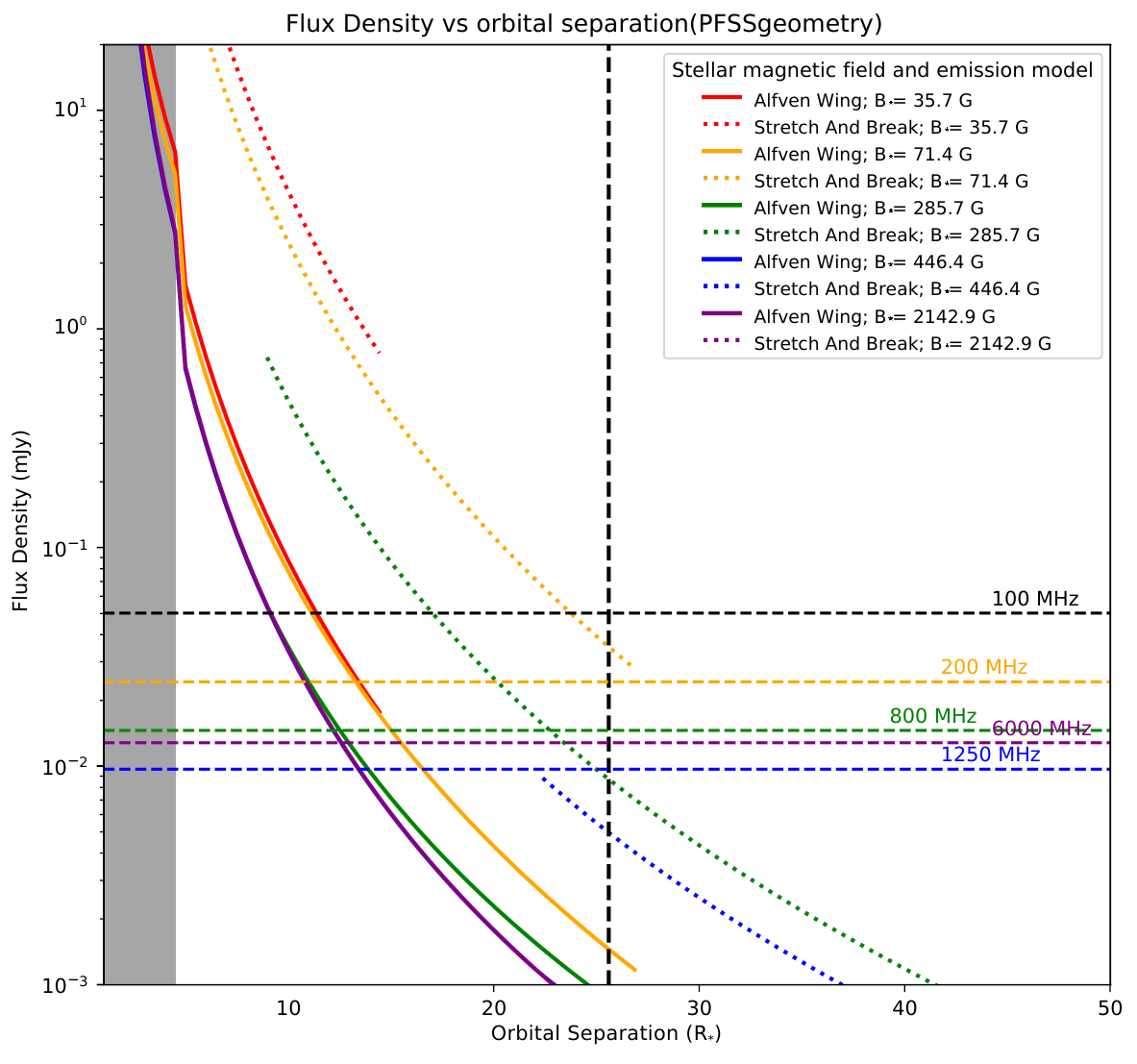}
\end{minipage}
\hfill
\begin{minipage}{0.35\textwidth}
\caption{Theoretical M-SPI radio flux density for an Earth-like exoplanet orbiting a mid-M-dwarf at 5\,pc for the Alfv\'en wing model (solid lines) and the stretch-and-break model (dashed-lines). The colours correspond to different stellar surface magnetic field values given in the legend. The horizontal dashed lines correspond to the SKA telescopes' sensitivity (version AA4) for a 100\,MHz bandwidth and 1\,hr integration. The vertical dashed line marks the semi-major axis of a 5\,day orbit.}
\label{fig:pfss_predict}
\end{minipage}
\end{figure}

Fig. \ref{fig:pfss_predict} shows the predicted radio power in the Alfv\'en wing and stretch-and-break models for a typical scenario: Earth-mass exoplanet with a field strength of $0.167\,{\rm G}$\footnote{Based on the scaling model of \citet{1993JGG....45...65S} for a tidally locked planet in a 5\,d orbit.} around a mid M-dwarf at 5\,pc. The radial evolution of the stellar field was calculated using the Potential Field Source Surface model \citep{1969SoPh....9..131A} where the area inside the `source-surface' is in gray. The stretch-and-break model shows no predicted flux density close to the star because $R_{\rm eff} < R_{\rm pl}$ and no planetary magnetosphere is formed. In addition, when the planet is far enough from the star, it would lie beyond the Alfv\'en surface and no star-side M-SPI emission would take place. The figure shows the rapid decline of M-SPI power with orbital separation. It also highlights the $\gtrsim 1$ order-of-magnitude difference between the power predicted by different models. Regardless, the plot demonstrates the anticipated impact of attaining $\mu$Jy-level sensitivities with the SKA.

\subsection{Ensemble flux density predictions}
The scaling laws from \S\ref{subsec:scaling_laws} in addition to a prescription for the stellar wind conditions at the location of the planet can be used to predict the radio flux density for both star-side and exoplanet-side emissions. The relevant input parameters of the stars are: (a) its large scale surface magnetic field, $B_\ast$, (b) its mass and radius, $M_\ast$ and $R_\ast$, (c) the base density of the stellar wind, $\rho_\ast$ and its rotation rate, $\Omega_\ast$, the coronal wind temperature, $T_\ast$ and optionally, the adiabatic index of the gas in the wind, $\gamma$. These determine the energy in the wind per unit time and surface area at any radial distance from the star. The necessary planetary parameters are: the surface magnetic field of the planet, $B_p$ (if it is magnetised), the radius of the planet, $R_p$ and the semi-major axis of the planet's orbit, $a_p$. With the stellar parameters in hand, the stellar wind density, $\rho$, temperature, $T$, velocity, $v$, and magnetic field, $B$, can be computed with the ideal MHD equations at the location of the planet. If the planet is magnetised then its effective size is that of its magnetopause where the planetary magnetic field is in pressure balance with the stellar wind's pressure. For an unmagnetized planet, the effective size is approximately that of the planet itself (most accurately, that of its ionosphere). With the stellar wind energy flux and the obstacle size, the prescriptions of \S\ref{subsec:scaling_laws} can be used to compute the radio power.

There are several practical problems in estimating the M-SPI radio power. The main ones are:
\begin{itemize}
    \item Often the stellar magnetic field and coronal base density, or equivalently, its wind mass loss rate are unknown as they are notoriously difficult to measure. Only a handful of dwarf stars have empirical mass-loss constraints \citep{2021ApJ...915...37W,2024NatAs...8..596K}. 
    \item The planet's magnetic field is essentially unknown. In fact a radio detection is the most viable method to measure this quantity.
    \item The Poynting flux to radio power conversion factor (i.e. the efficiency) is only empirical known for solar system planets. It is possible that the same efficiency may not hold at different energy scales at which the detectable exoplanets operate. 
\end{itemize}

Several authors have made predictions for the radio flux densities for both star-side and exoplanet-side emissions for known exoplanets-- see for e.g. \citet{2001Ap&SS.277..293Z,2007P&SS...55..598Z,2007A&A...475..359G,2018MNRAS.478.1763L,2018ApJ...854...72T,2023pre9.conf03092M}. The presence of multiple physical/empirical models (see \S\ref{subsec:scaling_laws}) and the above-mentioned practical issues means that the predictions of radio flux density can vary widely even for the same planet when done with different assumptions. Moreover the current pace of exoplanet discovery means that the list of `best targets for radio detection' itself is rapidly evolving. 
We have used here a prediction method, following \citet{2007A&A...475..359G,2023pre9.conf03092M,Mauduit2024}, to calculate the radio flux densities as an ensemble as described below.

\begin{itemize}
    \item {\em Planet parameters}: We started with the latest catalogue of `confirmed' exoplanets downloaded from \texttt{exoplanet.eu}. In cases where only the exoplanet mass, $M_p$ was available, we estimated its radius using 
    the prescription of \citet{Mauduit2024}, which improves upon earlier formulations (e.g. \citealt{2024A&A...686A.296M}).
    The magnetic field strength at the planetary surface, $B_p$, is derived from its magnetic dipole moment estimated 
    as the geometric mean of the models proposed by \citet{Busse1976}, \citet{Mizutani1992} (slow or moderate convection), and \citet{Sano1993}. An 
    alternate model based on \citet{Reiners2010} yields comparable results.
    \item {\em Stellar parameters}: We only considered entries where the stellar mass was known. We also rejected all pulsar and white dwarf hosts. 
    The large-scale stellar magnetic strength, $B_\ast$, is estimated from the database of Zeeman--Doppler Imaging measurements of \citet{Duchene2026}, which gathers magnetic field strengths and dipole field estimates for $\geq$2600 stars) and derives unmeasured magnetic fields by extrapolating them from the other available stellar parameters (mass, rotation period, effective temperature, etc.) with an accuracy $\sim \times 4$ at the $1\sigma$ level.
    If not measured, the stellar radius is then derived from the mass using a second-order polynomial fit established from the above database combined with the \texttt{exoplanet.eu} catalog: 
    \[
    \log_{10}(R_\ast/R_\odot) = 0.083 x^2 + 1.037 x + 0.018
    \]
    with $x = \log_{10}(M_\ast/M_\odot)$.
    This expression reproduces the empirical mass--radius relation, $R_\ast/R_\odot \propto (M_\ast/M_\odot)^\alpha$, with $\alpha \approx 0.8$ at low masses and $\alpha > 1$ for masses above about $0.5\,M_\odot$.
    The stellar rotation rate $\Omega_\ast$ is taken from the Simbad database (\url{https://simbad.u-strasbg.fr}), or estimated from the average of $v$ or $v . \sin i$ measurements listed in this database (with $median(1/|\sin i|) = (4/3)^{1/2}$ -- \citealt{2007A&A...475..359G}), or estimated from stellar age using the gyrochronological relations of \citet{Barnes2007}, \citet{Kounkel2023} or \citet{Newkirk1980} depending on the available stellar parameters.
    \item {\em Wind parameters}: We modelled the wind density, $\rho$, and velocity, $v$, using the one-dimensional (i.e. radial) isothermal Parker profile \citep{1965SSRv....4..666P}. The solution requires the following parameters to be fixed in addition to the stellar mass and radius. 
    (a) Isothermal wind temperature: For stars listed in the ROSAT catalogue of stars, we used the X-ray surface flux -- coronal temperature scaling of \citet{2015A&A...578A.129J}. For other stars we adopted the average solar value of $1.77\times 10^6\,$K. 
    (b) Wind mass-loss rate: The base wind density $\rho_\ast$ is derived from the stellar mass-loss rate, which depends on stellar age and is normalized to the solar mass-loss rate (at the star's age) per unit surface, as described in \citet{Griessmeier2007a} and \citet{Mauduit2024}.
    The stellar magnetic field in the wind is derived from the surface field $B_\star$ from \citep{Duchene2026}, assuming a smooth transition from a dipolar configuration ($B \propto (d/R_\ast)^{-3}$) for distances $d$ smaller than the `Alfv\'en point' $r_A$ where the wind velocity $v$ equals the Alfv\'en speed $v_A$, to a Parker spiral geometry with a radial component $B_r$ decreasing as $(d/r_A)^{-2}$ and an azimuthal component given by $B_\phi = B_r \Omega_\ast d / v$. The total magnetic field is then $B_{\mathrm{tot}} = (B_r^2 + B_\phi^2)^{1/2}$.
    This is a conservative assumption, as radial stretching of the field lines by the wind in the closed field region will lead to a drop off that is shallower than $(d/R_\ast)^{-3}$ which will increase the intercepted Poynting flux.
    \item {\em Energy calculation}: We assumed that the planetary magnetic pressure at radial distance (from the planets centre) falls off like a dipole (i.e. distance$^{-3}$) and estimated the effective planetary obstacle radius $R_{\rm mp}$ as the radial distance from the planet where its magnetic pressure equals the sum of the stellar wind magnetic and ram pressure. We then used a fixed conversion efficiency of $2\times 10^{-3}$ between the intercepted stellar wind Poynting flux and the emitted radio power from the star-side if the interaction was sub-Alfv\'enic \citep{2007P&SS...55..598Z}. Finally, we assumed an emission bandwidth equal to half the cyclotron frequency at the surface of the emitting body and the stellar parallax measured by {\em Gaia} to compute the expected flux density. 
    \item {\em Emission escape plus beaming}: There are several conditions that must be satisfied for the emission to reach the Earth: (i) the ECMI efficiency requires that the plasma frequency $f_{pe}$ is much smaller than the cyclotron frequency $f_{ce}$ (typically $f_{ce} > 10 f_{pe}$) at the source, (ii) $f_{pe}$ must remain lower than $f_{ce}$ along the propagation path, and (iii) emission must not be catastrophically absorbed near the source (but radiative transfer near the source requires knowledge of the plasma properties that we do not have). Here we simply required condition (i) to be fulfilled at the stellar surface, and the wind flow to be sub-Alfv\'enic in the stellar frame at the planet's orbit, in order to label the corresponding system `Emission likely'.
    We have no reliable information on the beaming geometry of most known exoplanet systems. As such, we omitted its effect on visibility save caution the reader that perhaps 10\% of radio-emitting exoplanetary systems are visible to any given observer.  
\end{itemize}

\begin{figure}
    \centering
    \includegraphics[width=\linewidth]{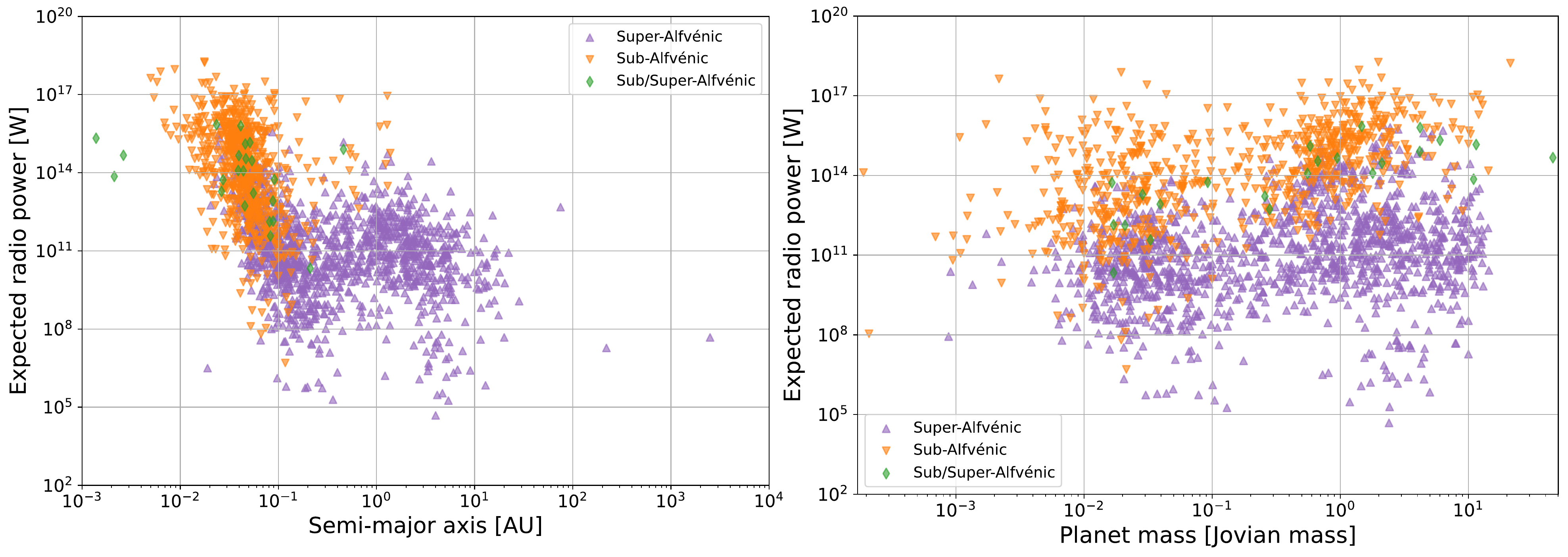}
    \caption{Intercepted Poynting flux times $2\times 10^{-3}$ (efficiency factor) for an ensemble of representative planet and star properties based on known exoplanets. Left and right panels show the flux against semi-major axis (of the planet's orbit) and the planet's mass, respectively. The points are colour coded by whether the interaction is expected to be sub- or super-Alf\'enic. In rare cases, the regime is sub-Alfénic in the star’s frame and super-Alfénic in the planet’s frame (due to its orbital speed).}
    \label{fig:ensemble_power}
\end{figure}

\begin{figure}
    \begin{minipage}{0.65\textwidth}
    \includegraphics[width=\linewidth]{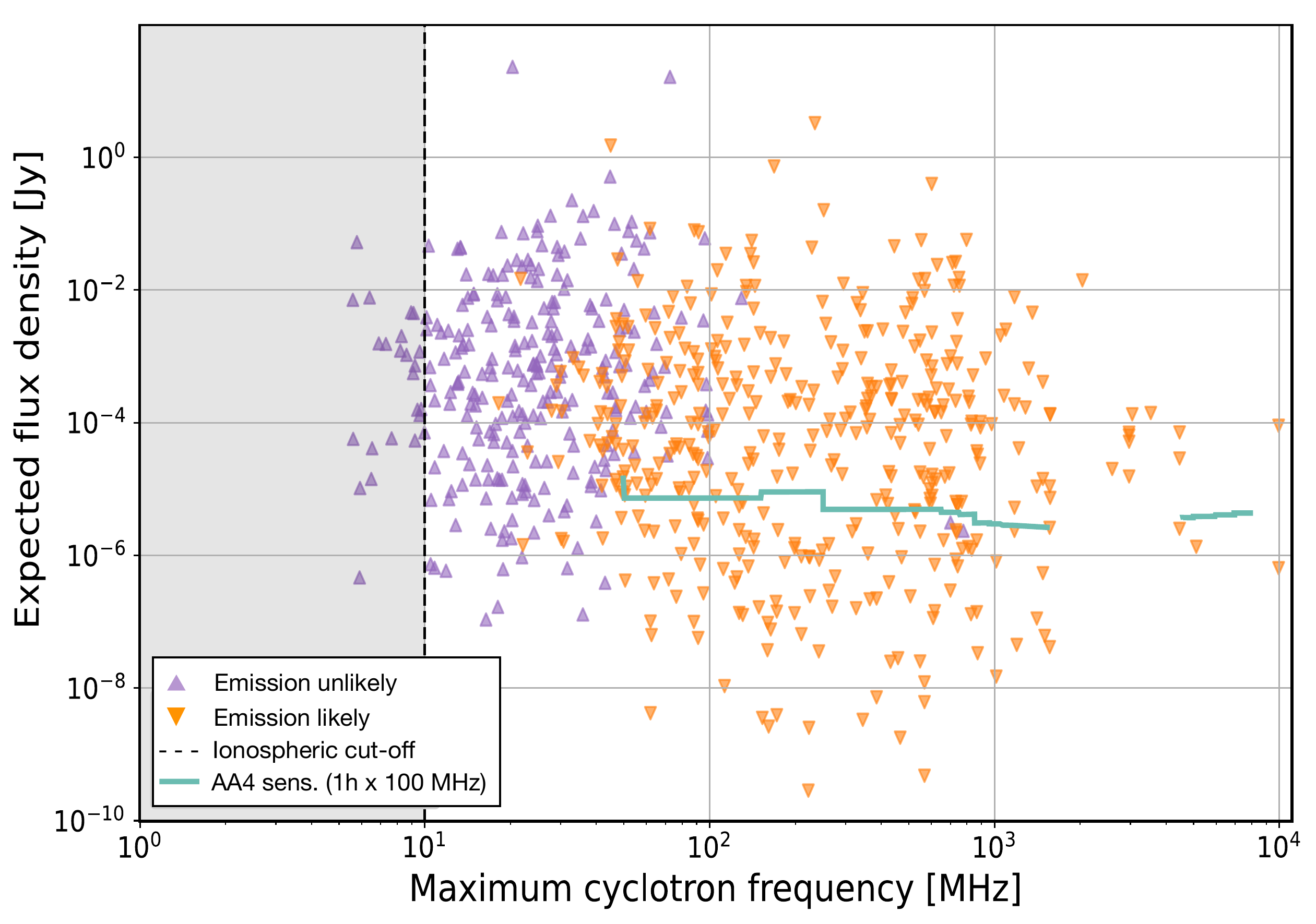}
    \end{minipage}
    \hfill
    \begin{minipage}{0.35\textwidth}
    \caption{Predicted stellar radio flux density (only for sub-Alfv\'enic interaction) as a function of the peak cyclotron frequency. The different colours denote cases where escape of radiation is likely (orange) and unlikely (magenta) respectively. The greenish-blue line marks the anticipated SKA telescopes' AA4 detection limit for a fixed 100\,MHz bandwidth and 1\,hr integration.}
    \label{fig:predicted_flux}
    \end{minipage}
\end{figure}

Figures \ref{fig:ensemble_power} and \ref{fig:predicted_flux} show the results of our ensemble simulations. The salient conclusion relevant for the SKA that we can draw from the ensemble simulations are as follows. 
\begin{itemize}
    \item On average, planets orbiting within 0.1\,AU have a high probability of being in the sub-Alfv\'enic zone; both star-side and planet-side emission from M-SPI is in principle detectable.
    \item The biggest determinant of the radio power appears to be the semi-major axis of the planet's orbit (or equivalently, its orbital period). The approximate dependence goes as $a^{-6}$ because of our (assumed) dipolar structure of the stellar field lines close to the star. Radio powers in the sub-Alfv\'enic zone could be expected to reach $~10^{17}\,{\rm W}$ and in the super-Alfv\'enic zone, we expect radio powers of up to $~10^{13-14}\,{\rm W}$. Due to the abundance of rocky planets in very short orbits, the dependence on planetary mass (or radius) is much weaker.
    \item A large area survey capable of detecting $\sim $mJy level sources in the SKA-Low band will be sensitive to a small fraction (a few percent) of exoplanets that happen to orbit low-mass stars (M-dwarfs mainly) in very short orbits ($a\ll 0.1\,{\rm AU}$). This is approximately the instrumental capabilities in the current `pre-SKA era'. If the current census of planet's in orbits $a<0.1\,{\rm AU}$ is assumed to be complete for stars within $\approx 50\,{\rm pc}$ then this represents 
    a few tens of detectable systems. Therefore fortuitous beaming is necessary for a secure M-SPI detection in the pre-SKA era. 
    \item SKA-Low AA4, being capable of $\sim 10\,\mu$Jy level sensitivity in modest integration times (1\,hr) will be sensitive to a much larger fraction (a few tens of percent) of radio-emitting exoplanets.  
    \item The star-side M-SPI radio emission could suffer ECMI quenching or radiation escape issues below $\sim 50\,{\rm MHz}$ which will not be an issue for observations with the SKA telescopes. 
    \item A large fraction of planet-side emission is likely to occur below the SKA-Low band making it unobservable. However it is plausible that planets above a Jovian mass emit in the SKA-low band and the most massive planets up to the brown dwarf limit could even be detectable up to 500\,MHz. Comprehensive predictions for the yield of exoplanet-side emissions are provided in \citet{Kavanagh01.2026.SKA}.
\end{itemize}

 \subsection{Visibility of SPI emission}
\label{subsec:tf_structures}
\begin{figure}
    \centering
    \includegraphics[width=\linewidth]{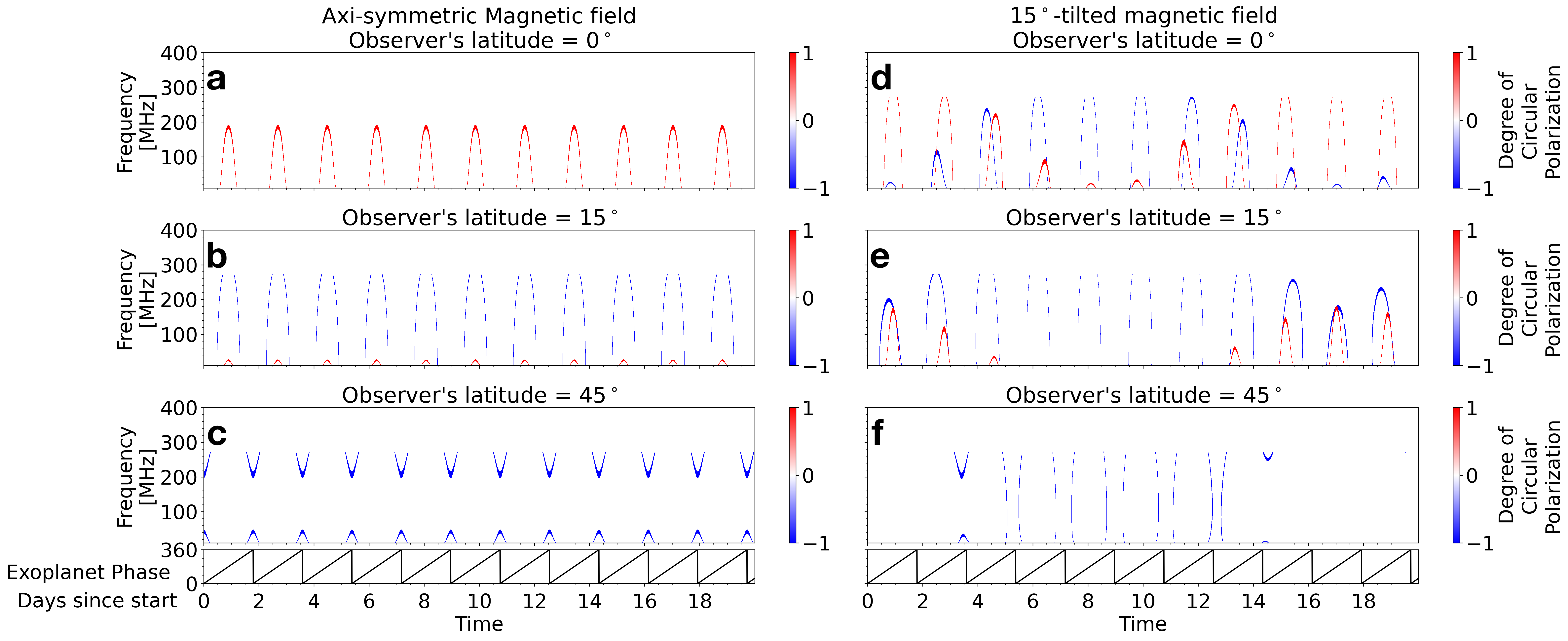}
    \caption{Simulations of SPI radio emissions (left column) for an axi-symmetric dipolar magnetic field and (right column) for a $15^\circ$-tilted dipolar magnetic field. The observer is fixed in the sky and located at (first row) $0^\circ$, (second row) $15^\circ$ and (third row) $45^\circ$ from the exoplanet orbital plane. The radio emissions are coloured by their degree of circular polarization: blue (-1) is for Right-Handed (i.e., emissions from the northern magnetic hemisphere) and red (+1) is for Left-Handed (i.e., emissions from the southern magnetic hemisphere). The phase of the exoplanet is indicated in the panels of the last row, with the exoplanet at opposition at $0^\circ$.}
    \label{fig:spi_emission_ExPRES}
\end{figure}

The beaming properties of ECMI emission not only determine which star--planet system's M-SPI signatures are visible to us but also produce unique time--frequency plane signatures that can be used to constrain the orbital and rotation geometry and the emitters magnetic-field topology \citep{2011A&A...531A..29H,2021JGRA..12629435L,2023MNRAS.524.6267K,2023pre9.conf03091L}. Indeed this characteristic periodic time--frequency patterns are the `smoking-gun' signature of star-side emission induced by M-SPI. 
In general, the visibility of SPI radio emissions in the time--frequency plane encodes information on the emitters magnetic field geometry (dipolar versus multipolar, axi-symmetric versus tilted/offset) and whether the angular momenta of rotation and orbital motion are aligned with the magnetic axes or not. As a simple example, consider the case of dipolar field with a low magnetic tilt in a system where the orbital and rotation axes are aligned. This is approximately true for the Jovian emission induced by Io. An observer located close to the magnetic equatorial plane will see emissions from both hemisphere (which have opposite sense of circular polarisation). An observer at higher magnetic latitude might only see radio emission from one magnetic hemisphere, and an observer at very high latitudes will see no emission at all \citep[see Figure 3d of][]{2021JGRA..12629435L}.

For other non-specific geometries, we must resort to numerical calculations. Here we use the ExPRES \citep[Exoplanetary and Planetary Radio Emission Simulator, ][]{2019A&A...627A..30L} code to simulate example visibilities to guide our observing strategy. This code has already been used to predict and reproduce radio emissions with success from the Jovian \citep{2008GeoRL..3513107H, 2017GeoRL..44.9225L, 2012P&SS...61...32C}, Kronian \citep{2008JGRA..11310213L, 2013JGRA..118.4817L} and Uranian \citep{2023pre9.conf03106L} systems, and also to make prediction for exoplanetary systems \citep{2011A&A...531A..29H}.

In Figure \ref{fig:spi_emission_ExPRES}, we show the results of six different simulations for two different magnetic field models (axisymmetric or $15^\circ$ tilted) and for three different positions of the observer (at $0^\circ$, $15^\circ$, $45^\circ$ of the exoplanet orbital plane). In each case, we used a $70$~G dipolar stellar magnetic field and a stellar rotation period of $2$~days. The exoplanet is located at $20$~stellar radii from the star, with an orbital period of $5$~days. The simulation run is $20$~days (indicated as the x-axis), to be sure to cover all possible geometries for this star-exoplanet system and periodicities.
The interaction between the exoplanet and the star is simulated using a loss-cone electron distribution function with a $20$~keV energy \citep[as observed for the Satellite-Planet Interaction at Jupiter, ][]{2023JGRA..12831985L}, with active sources on the magnetic field line connecting the exoplanet to the star. 

It has been shown that only emission beamed along the surface of a cone with an opening angle (from the local magnetic field line) larger than $70^\circ$ is visible: from theoretical calculations \citet{1986PhFl...29.2919P} showed the more the emission angle turns away from the perpendicular direction the weaker the emission is, reaching the background noise level for angles $<70^\circ$]; from comparison between ExPRES simulations and Juno measurements of the Io-Jupiter interaction radio emission, \citet{2017GeoRL..44.9225L} showed that only emission with a beaming angle $\ge 70^\circ \pm 5^\circ$ is visible; from in situ measurements in the source of the Satellite--Planet Interactions at Jupiter (Io, Europa and Ganymede) \citet{, 2023JGRA..12831985L} measured that the beaming angle is in the range [$70^\circ$ -- $90^\circ$]. For these reasons, only emission with a beaming angle $>60^\circ$ is display in Figure \ref{fig:spi_emission_ExPRES}.
Finally, only the frequency above $10$~ MHz, to mimic the ionospheric cutoff.

In Figure \ref{fig:spi_emission_ExPRES} we can see that in the case of an axi-symmetric magnetic field (\ref{fig:spi_emission_ExPRES}a-c)), the shape of the emissions is not evolving over time, while in the case of a $15^\circ$-tilted magnetic field, the shape of the radio emission is evolving over time. Indeed, for the tilted dipole the visibility of the emission is influenced by both the stellar rotation phase and the exoplanet's orbital phase.

For an observer located in the same plane as the exoplanet orbital plane (Figures \ref{fig:spi_emission_ExPRES}a,d), in the case of an axisymmetric magnetic field (Fig.  \ref{fig:spi_emission_ExPRES}a) both the star-exoplanet and exoplanetary system -- observer geometries are symmetrical, and both radio emissions from the southern (red) and northern magnetic hemispheres (blue, not visible) are overlapped. In the case of a $15^\circ$--tilted magnetic field (Figure \ref{fig:spi_emission_ExPRES}d), the visibility of emissions from one hemisphere or the other depends on the stellar longitude facing the exoplanet, and is therefore dependent on the geometry of the star-exoplanet-observer.

For an observer located at $15^\circ$ of the exoplanet orbital plane (Figures \ref{fig:spi_emission_ExPRES}b,e), one can see that only radio emission from the northern magnetic hemisphere (in blue) is visible, while for the case of a $15^\circ$--tilted magnetic field, some radio emissions from the southern magnetic hemisphere can still be seen for some star-exoplanet-observer geometry.

Emission at higher frequencies is visible for the cases where either the observer (Figures \ref{fig:spi_emission_ExPRES}b) or the magnetic field (Figures \ref{fig:spi_emission_ExPRES}d,e) are inclined by $15^\circ$, as the high-frequency sources are located closer to the planet, can only be beamed in the direction of the observer when the observer is far from the equator.

Finally, for an observer located at $45^\circ$ (Figures \ref{fig:spi_emission_ExPRES}c,f), most of the radio emission is no longer visible, more especially in the case of a $15^\circ$--tilted magnetic field. Indeed, the observer is located at too high--latitude, and most of the radio emissions are beamed away from the observer.

Note that we have not included the effects of refraction in the stellar corona plasma, which could alter the phases of visibility, particularly for emissions at higher frequencies (i.e. closer to the star). However, as long as the plasma parameters do not change with time, the periodicity in the visibility will endure. 
Although these simulations employ simplifying assumptions, they show examples of periodic time--frequency structures that must be expected in star-side M-SPI emission. The common property of these patterns are their `arch-like' pattern in the time-frequency plane and their periodic modulation either at the exoplanet orbital period (in case of an axi-symmetric stellar magnetic field) or at the beat period between the stellar rotation and the planet's orbit  \citep[i.e. the synodic period;][]{2023MNRAS.524.6267K, 2025arXiv250318733L}. These properties can be used as an efficient filter to find M-SPI candidates from amount radio-loud stars detected with the SKA telescopes. Additionally, auxiliary information about the star could be obtained by other means for promising candidates (e.g., Zeeman Doppler images of its magnetic field, spectroscopy/photometric measurements of its rotation period and inclination, etc.) to produce more realistic simulation \citep[see, e.g.,][]{2025Chebly}. Armed with this information and the forward modelling framework described here, an inversion of the data into primary system parameters becomes feasible \citep[see for e.g. ][]{2022MNRAS.514..675K}. 

The problem of inversion of the planet-side emission parameters is similar but much simpler. In this case the emission is periodic only at the rotation period of the planet that can in principle be distinguished from the stellar rotation period, planetary orbital period (or their synodic period therefore) by optical-band measurements of the stellar rotation period and/or the planetary orbital period. In this case the emission in the time--frequency plane still encodes information on the magnetic topology of the exoplanet and can be inverted using basic assumptions on the intrinsic beaming properties of ECMI emission \citep{2024A&A...692A..66K}. If successful this will perhaps prove the only technique to not just directly measure the magnetic field of an exoplanet but also determine its topology. 

\section{Observing strategy}
\label{sec:obs}
The observing strategy required to discover or confirm super- and sub-Alfv\'{e}nic M–SPI depends on several characteristics of the radio emission . Namely, the radio emission is:

\begin{itemize}
\item Beamed along the surface of a cone with a large opening angle ($\gtrsim70^{\circ}$);
\item Limited in frequency by the magnetic field strength of the emitting body ($\nu \propto B$);
\item Often highly circularly polarised ($>60\%$); and,
\item Modulated by either the orbital motion of the exoplanet or the rotation of the host body \citep{2011A&A...531A..29H,2023MNRAS.524.6267K,2025arXiv250318733L}. Usually this modulation results in a low ($\lesssim$10\%) duty cycle of emission (see Fig. \ref{fig:spi_emission_ExPRES}) but provides peculiar arch-like structures in the dynamic spectrum.
\end{itemize}

Since the radio emission is beamed, has a low duty cycle or is inherently stochastic in brightness, there is difficulty in \emph{a priori} identifying radio bright M-SPI systems despite stellar system parameters having ideal parameters for emission \citep{2025arXiv250318733L}. Moreover circularly polarised ECMI emission can be generated by stars due to their own magnetic activity, unrelated to any M-SPI. For example, despite almost all M\,dwarfs host a short period planet \citep{2015ApJ...807...45D}, \citet{2021NatAs...5.1233C} only identified $\approx$0.5\% of all M dwarf systems within 50\,pc as being radio-bright at 144\,MHz at $\sim $mJy sensitivity. Therefore, even in the era of the SKA telescopes, wide-field circularly polarised surveys will remain the workhorse for discovering M-SPI candidates.

Regardless, such wide-field surveys naturally fit with the capabilities of SKA-Low due to its large field-of-view. Conducting a survey of the entire southern sky with 1\,h integrations at 144\,MHz and a bandwidth of approximately 50\,MHz is expected to yield roughly 
$700 \pm 300$ detections in circular polarisation (extrapolating from the detection rates presented by \citet{2023A&A...670A.124C}). This estimate is based on a Stokes~V sensitivity threshold of at least $5\sigma$ (corresponding to 
$\gtrsim 140~\mu\mathrm{Jy}$), and assumes that the detection rate of variable 
circularly polarised sources in 1\,h observations matches that observed in 
8\,h epochs. The detection efficiency to the arch-like structures can be further improved by massively-parallel dynamic spectroscopy (i.e. coherently summing visibilities) in the direction of known stars as demonstrated by \citet{Tasse2026} with the LOFAR Two Metre Sky Survey data.  

While identifying circularly polarised emission is likely a necessary pre-requisite to identifying M-SPI candidates, confirming such a detection requires long-term monitoring of the candidates and detection of the characteristic period time--frequency structures such as the ones shown in Fig. \ref{fig:spi_emission_ExPRES}. Coverage of multiple exoplanet orbits may not be practical for a large number of M-SPI candidate systems detected as polarised radio sources in the SKA-telescopes' band. A practical approach to narrow down candidates, as also proposed by \citet{2024NatAs...8.1359C}), is to first secure a minimum of three radio detections of a stellar system at the same (putative) orbital phase -- ideally distinct from the star’s rotational phase. To date, no published claim of a detection of sub-Alfv\'{e}nic M-SPI in the radio-band has satisfied this criterion. A further way to build confidence in a detection would be the confirmation via traditional exoplanet detection methods \citep{2025arXiv250713783K}, such as radial velocities or transits, of the presence of an exoplanet in the orbit predicted from the radio emission modulation. 

For a super Alfv\'{e}nic M–SPI detection, detecting the modulation related to the rotation of the exoplanet would be convincing. This is because it will be difficult to disentangle whether the radio emission is emerging from the star or planet since the resolution of SKA-Low will be too low for even the nearest star systems. Alternatively, a detection of a low-frequency cut-off in the radio emission could also aid in supporting a claim of super Alfv\'{e}nic M–SPI detection, especially if the global magnetic field strength of the star is known from Zeeman Doppler Imaging.

Given the substantial investment of observatory time and resources required to confirm radio detections of SPI, it is essential that observational efforts focus on the most promising targets if not working from candidates known to be radio bright. To first order, the systems most likely to exhibit SPI-related radio emission can be identified based on planetary size, orbital separation, and system distance from Earth. Under these criteria, the leading candidates include 51~Pegasi, HIP~65~A, $\tau$~Bo\"{o}tis~A, 55~Cancri~A, and WASP-18 -- several of which have already been the focus of radio searches.  

However, many of these host stars may lack sufficiently strong magnetic fields to generate electron cyclotron maser (ECM) emission at frequencies $\gtrsim 100$\,MHz, where the SKA telescopes will be most sensitive. In addition, their complex magnetic topologies may prevent consistent beaming of emission toward Earth. Consequently, close-in planets orbiting M~dwarfs could represent more favourable targets, as these stars can possess kilogauss-strength, predominantly dipolar magnetic fields. Promising examples include GJ~367, GJ~436, GJ~1252, GJ~3253, GJ~625, YZ~Ceti, AU~Microscopii, and Proxima~Centauri \citep{2019MNRAS.488..633V}.

Taking these consideration, the predictions of \S\ref{subsec:scaling_laws} and \S\ref{subsec:tf_structures} together, we propose the following observational approach:
\begin{itemize}
    \item A wide-angle sky survey with SKA-Low reaching depths of $\sim 10^{-5}$\,Jy to identify circularly polarised stellar targets. Such a survey could be done within a wider collaboration due to its wide impact on nearly all radio-astronomical fields 
    \item A follow-up of the detected stars with known exoplanets to ascertain at least three detections at the anticipated orbital phase to form the nearly-confirmed M-SPI detections.
    \item A long-term multi-wavelength campaign of the nearly-confirmed candidates with the SKA telescopes to conclusively detect the characteristic time--frequency signatures such as the ones shown in Fig. \ref{fig:spi_emission_ExPRES}.
    \item A parallel observational effort targeting at least 10 known exoplanet systems where the availability of ancillary data allows us to realistically model the stellar wind properties leading to the prediction of high radio M-SPI flux  densities. These targets should be observed for at least 2 exoplanet orbits to check if they display recurring time--frequency signatures at the anticipated periodicity. 
\end{itemize}

\section{Conclusion \& Outlook}
While a large number of exoplanets have now been detected, their co-evolution pathways with host stars remains largely unknown. Of key importance to co-evolution for short period planets is the effect of magnetic star--planet interaction (M-SPI). Such interactions can lead to electrodynamic heating of the planets and erosion of their atmospheres. The interactions can also induce flares on the host stars modulating their magnetic activity evolution. The radio signatures of M-SPI are particularly interesting as they provide unique information on the rotation-orbital geometry, magnetic field strength and topology of stars and exoplanets. More importantly, radio detection of exoplanet-side M-SPI emissions are likely to be the only technique to directly measure the magnetic field strength and topology of exoplanets, and the SKA telescopes' prospects for such detections are discussed in \citet{Kavanagh01.2026.SKA}.  

Detecting the electromagnetic signatures of M-SPI is feasible but demands substantial telescope time investment because M-SPI signatures  must be `pulled out of' a sea of unrelated stochastic background signals with similar brightness and polarisation properties.
Thanks to the sustained investments in dedicated stellar and exoplanet observing telescopes in the optical band, M-SPI signatures have been robustly detected providing a proof of concept that electrodynamic processes and magnetic fields around exoplanets is now within observational reach. Here we have made a case for the SKA telescopes with their unprecedented sensitivity (and survey speed) to join the fray to realise the promise of radio M-SPI studies. 

Based on our current best understanding of the expected phenomenology of M-SPI radio emission (described in \S\ref{subsec:scaling_laws} and \S\ref{subsec:tf_structures}), we have outlined a two-tier observational strategy in \S\ref{sec:obs} that combines relatively shallow ($\sim 10^{-5}\,{\rm Jy}$) wide-angle survey with long term (multiple exoplanet orbits) radio monitoring of promising targets. The former can be conducted as a community-supported survey. The latter will require substantial resources or the order of few hundred hours of telescope time per target. We end by stressing the importance of such long-term monitoring in the radio band: the detection of M-SPI in the optical flare arrival time distribution by \citet{2025Natur.643..645I} came after nearly 70\,days of space telescope monitoring, whereas the spectroscopic detection of M-SPI in chromospheric Ca lines in GJ\,436 came after observations spanning two decades \citep{Revilla2026}. Similar efforts are necessary in the radio band to realise the high-impact scientific potential of radio observations to stellar and exoplanet science.

\section*{Acknowledgements}
This research made use of data obtained from or tools provided by the portal exoplanet.eu of The Extrasolar Planets Encyclopaedia, and the VizieR catalogue access tool, CDS, Strasbourg, France. HKV acknowledges funding from the European Research Council under the European Union’s Horizon Europe programme (grant number 101042416 STORMCHASER). HKV, RDK, and SB acknowledge funding from the Dutch research council (NWO) under the talent programme (Vidi grant VI.Vidi.203.093). A.S acknowledges funding from the PLATO/CNES grant at CEA/IRFU/DAp, and the European Research Council project ExoMagnets (grant agreement no. 101125367). LPM, MPT and PJA acknowledge financial support from the Agencia Estatal de Investigaci\'on (AEI/10.13039/501100011033) of the MICIU and the ERDF ``A way of making Europe'' through projects PID2023-147883NB-C21, PID2022-137241NB-C43,  and the Centre of Excellence ``Severo Ochoa'' award to the Instituto de Astrof\'isica de Andaluc\'ia (CEX2021-001131-S). PZ, EM and ND acknowledge funding from the European Research Council (ERC) under the European Union’s Horizon 2020 research and innovation programme (grant agreement N$^\circ$101020459 -- Exoradio).\\
{\em Software used}: \texttt{python3.10}, \texttt{numpy}, \texttt{matplotlib}, \texttt{astropy}, \texttt{astroquery}.\\

\bibliographystyle{abbrvnat-maxbibnames4}
\bibliography{references} 

\end{document}